\documentclass[12pt]{article}
\usepackage{amssymb}
\usepackage{epsfig}
\usepackage{amsmath}
\begin{document}

\rightline{hep-th/0403170}

\bigskip\bigskip

\begin{center} 
{\Large \bf Brane-antibrane systems and \\ the thermal life of neutral
black holes}
\end{center} 

\bigskip\bigskip
  
\renewcommand{\thefootnote}{\fnsymbol{footnote}} 
\centerline{\bf
Omid~Saremi\footnote{omidsar@physics.utoronto.ca} and
Amanda~W.~Peet\footnote{peet@physics.utoronto.ca}}
\bigskip
\centerline{\it Department of Physics,}
\centerline{\it University of Toronto,}
\centerline{\it 60 St. George Street,}
\centerline{\it Toronto, Ontario,}
\centerline{\it Canada M5S 1A7.}
   
\setcounter{footnote}{0}   
\renewcommand{\thefootnote}{\arabic{footnote}}   

\bigskip\bigskip

\abstract{A brane-antibrane model for the entropy of neutral black
branes is developed, following on from the work of Danielsson, Guijosa
and Kruczenski [1].  The model involves equal numbers of D$p$-branes
and anti-D$p$-branes, and arbitrary angular momenta, and covers the
cases $p=0,1,2,3,4$.  The thermodynamic entropy is reproduced by the
strongly coupled field theory, up to a power of two.  The
strong-coupling physics of the $p=0$ case is further developed
numerically, using techniques of Kabat, Lifschytz et al. [2,3], in the
context of a toy model containing the tachyon and the bosonic degrees
of freedom of the D0-brane and anti-D0-brane quantum mechanics.
Preliminary numerical results show that strong-coupling
finite-temperature stabilization of the tachyon is possible, in this
context.}

\vfill

\noindent 16 March 2004.

\setcounter{page}{0}
\newpage
\section{Introduction}
The drive to explain the thermodynamic entropy of black holes and
black branes via the statistical mechanics of microscopic degrees of
freedom has been a central preoccupation of string theorists and other
gravitational theorists for at least two and a half decades.  String
theory has provided significant progress on this front in the last
eight years.  The celebrated 1996 success of Strominger and Vafa in
computing the entropy of particular $D=5$ supersymmetric (BPS) black
holes occurring in low-energy string theory, by using a conformal
field theory appropriate to the microscopic physics of strings and
D-branes, even raised the profile of string theory itself in a
significant way.  Further successes followed; the thermodynamic
entropy of BPS and near-BPS black holes in various dimensions was
reproduced successfully, in some cases even beyond leading order in
macroscopic quantum numbers such as mass, charge, and angular momenta.
A microscopic string theoretic accounting of the entropy of neutral
black holes and black branes remains more elusive, however. One reason
is that neutral black holes and branes are literally as far from BPS
as possible, meaning that supersymmetry may provide no help at all in
the endeavour.

BFSS matrix theory \cite{BFSS} is a conjectured relationship between
quantum $D=11$ M theory in light-front frame and the supersymmetric
quantum mechanics of a large number $N$ of D0-branes.  This theory was
recruited, e.g. in \cite{KlebanovSusskind}, to help explain the
entropy of near-BPS black holes in dimensions $D=10-p$, and also in
attempts to explain the entropy of the neutral cases as well.  Reviews
of this story include \cite{peetCQGreview} (see references therein).
Another interesting proposal was in \cite{Englert}.  Here, we simply
record one major common feature of these various matrix theory
approaches to counting the entropy of neutral black holes, which is
relevant to our work here.  Namely, the need for an infinite boost in
the 11th dimension $x^{11}$ in order to relate the neutral black holes
of interest to nearly-BPS systems whose entropy can be computed
microscopically using D-brane field theory.  In particular, this
infinite boost in the 11th dimension eliminates anti-D0-branes from
the picture.  The approach that we will take here will be different.
Other interesting approaches to neutral black hole entropy from quite
different perspectives include \cite{dasetal} and references therein.

Gravity/gauge correspondences are specific relationships between open
string and closed string degrees of freedom.  For the case of sixteen
supercharges \cite{itzhaki}, they relate the degrees of freedom of
supersymmetric Yang-Mills theories to closed string theory on
near-horizon D$p$-brane spacetimes.  Subsequent string theoretic
developments of interest for our work here include tachyon
condensation in brane-antibrane systems.  The paper which sparked our
specific interest was \cite{Danielsson}, in which an interesting
proposal was made for a microscopic explanation of the entropy of
neutral black D3, M2 and M5 (non-dilatonic) branes in terms of a
system of strongly coupled branes and antibranes.

In section 2, we review features of earlier analytic work, upon which
we build here.  Section 3 contains the bulk of our analytic
observations.  We study the entropy of neutral black D$p$-brane
spacetimes by using a microscopic model with an equal number of
strongly coupled D$p$-branes and anti-D$p$-branes, with arbitrary
angular momenta turned on.  Our main result is that the supergravity
entropy density can be reproduced by strongly coupled field theory, up
to a power of two.  Section 4 contains an exposition of preliminary
numerical work on the $p=0$ system.  Our primary goal in the numerical
part of our work is to try to test some assumptions of the
brane-antibrane model, by doing direct strong-coupling simulation of
the $p=0$ system, adapting methods of \cite{kabat1,kabat2}.  We use a
bosonic toy model to do this analysis, and find preliminary evidence
that tachyon stabilization may indeed occur in the fashion expected
from the work of \cite{Danielsson}.  We are not yet able, however, to
be conclusive about the supersymmetric case.

While this paper was in preparation, the work of \cite{Guijosa}
appeared, which has some overlap with section 3, for the case $p=3$
with angular momentum.

\section{Analytic ingredients from prior work}

\subsection{Gravity/gauge correspondences for general $p$}

For a system of $N$ D$p$-branes, it is possible\footnote{For $p\leq
4$, which are the cases on which we will concentrate} to take a clean
low-energy limit such that the open strings with endpoints on the
D$p$-branes decouple from the closed strings in the bulk.  This
observation led, of course, to the celebrated AdS/CFT correspondence
and the non-conformal open/closed-string correspondences of Itzhaki,
Maldacena, Sonnenschein and Yankielowicz \cite{itzhaki}.  The theory
on the branes is supersymmetric Yang-Mills theory (SYM) with sixteen
supercharges.  The 't Hooft coupling for the SYM theory is
\begin{equation}
g_{\rm YM}^2N\equiv(2\pi)^{p-2}g_s\ell_s^{p-3}\ N \,,
\end{equation}
where $g_s$ is the string coupling and
$\ell_s\equiv\sqrt{\alpha^\prime}$ is the string length.

These open/closed string correspondences tell us that we can hope to
understand the gravitational fields of D$p$-branes via the dual SYM
theory.  For $p\not=3$ the fields describing the $D=10$ geometry vary
radially in the spacetime, and on the SYM side the coupling is
dimensionful, yielding breakdown of the weak-coupling description
either in the UV or the IR.  In other words, there are limits to the
validity of both descriptions.  A dimensionless control variable is
given by $g_{\rm eff}^2\equiv g_{\rm YM}^2N U^{p-3}$; on the
supergravity side, $U\equiv r/\ell_s^2$ is the radial isotropic
coordinate in energy units.  The requirements \cite{itzhaki} for the
supergravity geometry in $D=10$ to remain valid are
\begin{equation}
1\ll (g_{\rm YM}^2N U^{p-3})  \ll N^{\frac{4}{(7-p)}} \,.
\end{equation}
At the left-hand end, $\alpha'$ corrections become important and the
SYM theory takes over from the $D=10$ geometry as the weakly-coupled
description; at the right-hand end, strong coupling (dilaton) ensues
and it is necessary (for large-$N$) to turn to the S-dual supergravity
geometry.

It was earlier noted in \cite{KlebanovTseytlin} that thermodynamic
properties of near-BPS D$p$-brane (and M-brane) geometries could be
written in a way that is suggestive of a field theory interpretation.
The main focus of that paper was the non-dilatonic branes, for which
the energy density above extremality $\Delta m$ and entropy density
$s$ were written in a way reminiscent of gases of weakly interacting
massless particles in $d=p+1$.  For the dilatonic cases, however, the
expressions do not yield recognizable weak-coupling results. The
physical interpretation of this fact is that the strongly coupled
physics of the $p\not=3$ SYM theories gives rise to nontrivial
dependence of the entropy and the energy above extremality on the
temperature,
\begin{equation}\label{klebtseysimple}
\Delta m(T) \propto T^{\frac{2(7-p)}{(5-p)}} \,,\qquad s(T)
\propto T^{\frac{(9-p)}{(5-p)}} \,,
\end{equation}
where $T$ is the Hawking temperature of the geometry.

In the $p=0$ case, checking aspects of the open/closed-string
correspondence conjectures is potentially feasible.  The reason is of
course that the SYM theory in this case is actually matrix {\em
quantum mechanics} -- albeit with sixteen supercharges.  The numerical
investigations of \cite{kabat1,kabat2} aimed to check the $p=0$
correspondence explicitly, by finding the entropy of the system of
{\em strongly coupled} $N$ D0-branes and comparing it to the entropy
of the near extremal D0-brane supergravity background.  Results
obtained showed that the 9/5 power in the expression for the entropy
in (\ref{klebtseysimple}) was indeed approximately reproduced in the
numerical approach.  This striking result, in combination with the
work given a lightning review in the next subsection, provided the
essential motivation for our work.

\subsection{Brane and antibranes at finite temperature and the entropy
of black branes}
Motivated by an observation of Horowitz, Maldacena, and
Strominger~\cite{Maldacena1}, Danielsson, Guijosa and Kruczenski
\cite{Danielsson} argued that, starting with a brane-antibrane system
at zero temperature, turning on a finite temperature could lead to
reappearance of open string modes.  This physics was argued to ensue
at strong open-string coupling $g_s N$ but weak closed-string coupling
$g_s$, as would be appropriate to a model with validity in the black
brane regime.  Note that, to have this reappearance of the open string
modes at temperatures {\em below} the Hagedorn temperature, going to
the regime of strong open-string coupling was crucial. In alternative
language, the tachyonic mode becomes stabilized by finite-temperature
and strong coupling physics. Based on these observations, the authors
of \cite{Danielsson} formulated a microscopic brane-antibrane model
for the entropy of non-dilatonic black branes: the D3, M2 and M5
cases.

The model has two SYM theories on worldvolumes, one for the branes and
the other for the anti-branes.  The tachyon, argued to be stabilized
by finite-temperature and strong coupling effects, does not contribute
to the total energy of the black hole, because it sits almost on top
of the tachyon potential. It is also argued that the tachyon does not
contribute to the total entropy, because the tachyon gets a large
thermal mass.

The black holes of interest to us are neutral.  Since the black
3-brane in $D=10$ (or, equivalently, the black hole in $D=7$) is
neutral, it is taken to be modelled by the gas on a number $N$ of
D3-branes, the gas on an identical number $N$ ${\overline{\rm
D3}}$-branes, and the tachyon dynamics.  Since the branes and
anti-branes appear in equal numbers, it makes sense to assume that the
corresponding temperatures are equal.  Using these assumptions, for
the strongly coupled field theory side,
\begin{eqnarray}
m_{\rm FT} = 2N\tau_3+a\frac{\pi^2}{8}N^2 T^4, \nonumber\\ 
s_{\rm FT}=a\frac{\pi^2}{6}N^2T^3 \,,
\end{eqnarray}
where $\tau_3$ is the D3-brane tension and $a$ is a constant known
\cite{peet1} from AdS/CFT to be 6 (not 8) when the SYM theory in
question is strongly coupled.  Note that the quantities of
thermodynamic interest written here are specific, i.e.  mass and
entropy {\em densities}.  Rearranging to eliminate $T$ gives
\begin{equation}
s_{\rm FT} = a(\pi^2/6) N^2 \left(
{\frac{m_{\rm FT} - 2N\tau_3}{a(\pi^2/8)N^2}}
\right)^{3/4} \,.
\end{equation}

It is important to emphasize that this analysis is done in the {\em
microcanonical} ensemble, where the total mass and charge of the black
hole, as well as any angular momenta present, are kept fixed.  Working
in the microcanonical ensemble prevents thermal fluctuations from
creating an infinite number of D3-${\overline{\rm D3}}$ pairs.
Working in canonical picture, by contrast, would allow the system to
catastrophically create an infinite number of pairs, by extracting an
arbitrary amount of energy from the reservoir, as it is a
thermodynamically favourable state. In other words, one indefinitely
spends energy to make more entropy, which makes the thermal ensemble
destabilized.

The essential innovation in \cite{Danielsson} is that the total number
of branes is regarded as an independent variable, which is determined
thermodynamically.
The next step in the analysis, then, is to maximize the entropy with
respect to $N$.  The value of $N$ so obtained is related to the mass
density as
\begin{equation}
m_{\rm FT} = 5 N \tau_3 \,.
\end{equation}

To proceed further, it is necessary to decide what to do with the
masses.  One assumption is to take $m_{\rm FT}=m_{\rm SG}$.  Using
this, and substituting back for the optimized value of $N$ leads to
the final expression for the entropy from the strongly coupled field
theory side,
\begin{equation}
s_{\rm
FT}=a^{\frac{1}{4}}2^{\frac{5}{4}}3^{-\frac{1}{4}}5^{-\frac{5}{4}}
\pi^{\frac{1}{4}}\sqrt{\kappa} m_{\rm FT}^{\frac{5}{4}} \,,
\end{equation}
where $\kappa={\sqrt{\pi}}/{\tau_3}$. One the other hand, entropy of a
black 3-brane is
\begin{equation}
s_{\rm SG}=2^{\frac{9}{4}}5^{-\frac{5}{4}}\pi^{\frac{1}{4}}
\sqrt{\kappa}m_{\rm SG}^{\frac{5}{4}} \,.
\end{equation}
Identifying $m_{\rm SG}=m_{\rm FT}$ and $a=6$ gives
\begin{equation}
s_{\rm SG}= 2^{3/4} s_{\rm FT} \, .
\end{equation}
This was the result of \cite{Danielsson}.  It shows that the entropy
scaling from the brane-antibrane model is correct.  It is worth noting
here that this scaling agreement is nontrivial: it does {\em not} come
about through simple dimensional analysis (and counting powers of
$N$).

The overall coefficient misses, by a factor close to unity.  It is
perhaps not surprising, though, that the agreement, while close, is
not exact.  Reasons for this may include the fact that there is no
clean decoupling between open string and closed string degrees of
freedom.  These effects not taken into account in the brane-antibrane
model may indeed play a role, but they cannot be major effects for
quantities like the entropy, because the scaling comes out correct.

\section{D$p$ and ${\overline{\rm\bf D}p}$-branes
with angular momenta}
In order to test the idea of the brane-antibrane model for neutral
black hole entropy further, it is interesting to consider the cases
other than $p=3$, and to add angular momentum.  This will be the main
focus of the analytic part of our work.

It is convenient to begin by reviewing some salient properties of
rotating black D$p$-branes for various $p$.  We are particularly
interested in the $p=0$ story, since that is the case of the
brane-antibrane model which we plan to test numerically, starting with
the preliminary investigations of section 4.

\subsection{Supergravity side}

Black $p$-branes in $D=10$ are of course equivalent to $D\equiv 10-p$
dimensional black holes.  We will not need the precise form of the
supergravity fields; what is important for us here is the relationship
between various physical parameters including tension, charge, angular
momenta, and horizon radius.  Using for example \cite{HarmarkObers},
and converting to quantities which are specific (per unit volume), we
have for the $D=10$ black branes
\begin{eqnarray}
j_i &=& {\frac{2}{(7-p)}}{\frac{\alpha_p}{16\pi G}}
r_0^{7-p}\ell_i {\frac{1}{\sqrt{1-\zeta^2}}} \,,\nonumber\\
T_H &=& {\frac{(7-p)-2\kappa}{4\pi r_H}} {\sqrt{1-\zeta^2}}
\quad {\rm{where}}\quad \kappa = \sum_{i=1}^n
{\frac{\ell_i^2}{\ell_i^2+r_H^2}} \,,\nonumber\\
m &=& {\frac{\alpha_p}{16\pi G}} r_0^{7-p} \left[
{\frac{1}{(7-p)}} + {\frac{1}{1-\zeta^2}} \right] \,,\nonumber\\
s &=& {\frac{4\pi}{(7-p)}} {\frac{\alpha_p}{16\pi G}} r_0^{7-p}
r_H {\frac{1}{\sqrt{1-\zeta^2}}} \,,\nonumber\\
q &=& {\frac{\alpha_p}{16\pi G}} r_0^{7-p} \left[
{\frac{\zeta}{1-\zeta^2}} \right] \,,
\end{eqnarray}
where $\zeta$ is the boost rapidity parameter while $r_0,\ell_i$ are
the original parameters used in creating the solutions, while the
Newton constant is given by $16\pi G = (2\pi)^{7} g_s^2 \ell_s^8$.  We
also use the shorthand $ \alpha_p \equiv (7-p)\Omega_{8-p}$.  Values
of $\alpha_p$ are tabulated here for convenience
\begin{center}
\begin{tabular}{|c|c|c|c|c|c|c|c|}\hline
$p$ & 0 & 1 & 2 & 3 & 4 & 5 & 6
\cr\hline
$\alpha_p$ & $7\pi^4/3$ & $32\pi^3/5$ & $5\pi^3$ & $32\pi^2/3$ &
$6\pi^2$ & $8\pi$ & $2\pi$
\cr\hline
\end{tabular}
\end{center}
The parameter $r_H$ is the horizon radius given by
\begin{equation}\label{r0rH}
r_H^{7-p} \prod_{i=1}^{n}\left( 1 + {\frac{\ell_i^2}{r_H^2}} \right) -
r_0^{7-p} = 0 \,.
\end{equation}
It is straightforward to massage the expressions to write the entropy
density $s$ in terms of the energy density above extremality $\Delta
m$.  To do that, it is convenient use (\ref{r0rH}) to obtain $r_0$ in
terms of $r_H$ and $\ell_i$.  This is simple if we make the
definitions
\begin{equation}
\rho \equiv {\frac{r_H}{r_0}} \ \,, \quad
\lambda_i \equiv {\frac{\ell_i}{r_H}} \,.
\end{equation}
Then we have
\begin{equation}\label{rhoeqn}
\rho(\lambda_i) \equiv {\frac{r_H}{r_0}} =
\left[ \prod_{i=1}^n
(1+\lambda_i^2) \right]^{-{\frac{1}{(7-p)}}} \,.
\end{equation}
The energy density above extremality $\Delta m \equiv m-q$ is given by
\begin{equation}\label{deltaE}
\Delta m = {\frac{\alpha_p}{16\pi G}} r_0^{7-p} \left[
{\frac{1}{(1+\zeta)}} + {\frac{1}{(7-p)}} \right] \,.
\end{equation}
We can also write the Hawking temperature as
\begin{equation}
T_H = \sqrt{1-\zeta^2} \, {\frac{(7-p)}{4\pi r_0}}
{\frac{k_p(\lambda_i)}{\rho(\lambda_i)}} \,,
\end{equation}
where $\rho(\lambda_i)$ is given by (\ref{rhoeqn}) and we use the
additional shorthand
\begin{equation}
k_p(\lambda_i) \equiv 1 - {\frac{2}{(7-p)}}\sum_{i=1}^n
{\frac{\lambda_i^2}{(1+\lambda_i^2) }} \,.
\end{equation}
In addition, the number of D$p$-branes can be written as
\begin{equation}\label{Npzeta}
N_p = {\frac{\alpha_p}{16\pi G}} {\frac{1}{\tau_p}} r_0^{7-p}
{\frac{\zeta}{1-\zeta^2}} \,,
\end{equation}
where
\begin{equation}
\tau_p = {\frac{1}{g_s(2\pi)^p\ell_s^{p+1}}}
\end{equation}
is the D$p$-brane tension.

Our main interest is to explain the entropy of neutral black branes in
$D=10$, or equivalently, neutral black holes in $D=10-p$.  We
therefore record the expressions here that we aim to reproduce using
the strongly coupled brane-antibrane field theory model.

In the case of a neutral black hole, we have $\Delta m = m_{\rm SG}$.
Defining
\begin{equation}
\delta_p \equiv {\frac{(8-p)}{(7-p)}} \,,
\end{equation}
we have for the energy density
\begin{equation}
m_{\rm SG} = {\frac{\alpha_p}{16\pi G}} r_0^{7-p} \delta_p
\end{equation}
and for the Hawking temperature
\begin{equation}
T_H = {\frac{(7-p)}{4\pi r_0}}
{\frac{k_p(\lambda_i)}{\rho(\lambda_i)}} \,.
\end{equation}
Therefore,
\begin{equation}
m_{\rm SG} (T_H) = {\frac{\alpha_p}{16\pi G}} \delta_p \left(
{\frac{(7-p)}{4\pi}} {\frac{k_p(\lambda_i)}{\rho(\lambda_i)}}
{\frac{1}{T_H}} \right)^{7-p} \,,
\end{equation}
while for the entropy
\begin{equation}
s_{\rm SG} (T_H) = {\frac{4\pi}{(7-p)}} {\frac{\alpha_p}{16\pi G}}
\rho(\lambda_i) \left( {\frac{(7-p)}{4\pi}}
{\frac{k_p(\lambda_i)}{\rho(\lambda_i)}} {\frac{1}{T_H}} \right)^{8-p}
\,.
\end{equation}
Defining
\begin{equation}
a_p \equiv {\frac{(7-p)}{2}} \delta_p^{\delta_p}
\left({\frac{\alpha_p}{16\pi}}\right)^{\delta_p-1} \,,
\end{equation} 
we have that
\begin{equation}
s_{\rm SG}(m_{\rm SG}) = {\frac{2\pi}{Ga_p}} \rho(\lambda_i)
(Gm_{\rm SG})^{\delta_p} \,.
\end{equation}

We may now ask how to unfurl the dependence of $\rho(\lambda_i)$ on
$m_{\rm SG}$ and the rotation parameters $j_i$.  The simplest way to
proceed is to recognize that
\begin{equation}
{\frac{\lambda_i}{2\pi}} = {\frac{j_i}{s}} \,.
\end{equation}
We therefore have the equation
\begin{eqnarray}\label{sugralambda}
\lambda_i \left[\prod_{i=1}^n (1+\lambda_i^2)
\right]^{-{\frac{1}{(7-p)}}} &=& a_p {\frac{j_i}{G^{\delta_p-1}m_{\rm
SG}^{\delta_p}}} \nonumber\\
&=& {\frac{j_i}{R_{\rm SG}(m_{\rm SG},G)}} \,,
\end{eqnarray}
where
\begin{equation}
R_{\rm SG} \equiv {\frac{1}{a_p}} G^{\delta_p-1} m_{\rm SG}^{\delta_p}
\,.
\end{equation}
Then, finally, we have
\begin{equation}\label{sSGfnofrho}
s_{\rm SG} = 2\pi \rho(R_{\rm SG}(m_{\rm SG},G),j_i) \, R_{\rm
SG}(m_{\rm SG},G) \,.
\end{equation}

We want to invert (\ref{sugralambda}) to unfurl the dependence of
$\lambda_i$ on $m_{\rm SG}, G, j_i$.  When no angular momenta are
present, we have of course that $\rho=1$.  In general, however, the
equation for $\lambda_i$ is a polynomial of degree $(7-p)$ in
$\lambda$, which is a quintic or worse for $p\leq 2$.  For now, we do
not concern ourselves with whether we can actually solve for $\rho$;
we just leave the dependence of $\rho$ on $(m_{\rm SG},G,j_i)$
implicit.  We also note that for $p=3$ and one angular momentum
parameter $j_1$, we can actually solve to find
\begin{equation}
\lambda_1 (p=3) =
\sqrt{2\sqrt{\chi}\left(\sqrt{\chi}+\sqrt{1+\chi}\right)}
\end{equation}
and
\begin{equation}
\rho(p=3) = {\frac{1}{\sqrt{\sqrt{\chi}+\sqrt{1+\chi}}}} \,,
\end{equation}
where
\begin{equation}
\chi = 2^{-5} 3^{-1} 5^5 \pi {\frac{j_1^4}{G\,m_{\rm SG}^5}} \,.
\end{equation}
%

\subsection{Near-extremal: the field theory side}
In this subsection, we take the near-extremal limit of various
supergravity formul\ae\ to tell us how the strongly coupled field
theory quantities should behave on the branes and anti-branes, using
IMSY duality.

We should mention that in some respects our analysis here is quite
similar to \cite{KlebanovTseytlin}.  There are two major differences,
however. The first is that we are interested in the effect of turning
on angular momenta $j_i$.  Our results must of course reduce to those
of \cite{KlebanovTseytlin} upon taking $j_i=0$.  The second is that we
plan to use the information about the strongly coupled field theory,
obtained via weakly coupled supergravity, using the techniques of
\cite{Danielsson}.

Using the relation (\ref{Npzeta}), we can\footnote{The constant $c_p$
familiar from the TASI-99 lectures of one of us is related to the
quantities used here by $c_p g_s \ell_s^{7-p} = 16\pi G/\alpha_p$.}
express $r_0$ in terms of $N_p$
\begin{equation}\label{r0Npne}
r_0^{7-p} = 2(1-\zeta) \, N_p \tau_p {\frac{16\pi G}{\alpha_p}} \,,
\end{equation}
(Of course, this $r_0$ is appropriate to the near-extremal geometry,
and is not the same as the $r_0$ at the end of the previous
subsection, which refers to the horizon radius of the {\em neutral}
spacetime.) Equivalently,
\begin{equation}
2(1-\zeta) = \left({\frac{r_0}{\ell_s}}\right)^{7-p}
{\frac{(2\pi\alpha_p)(2\pi)^{2(p-5)}}{\ell_s^{3-p}(g_{\rm YM}^2N)}}
\,.
\end{equation}
The energy density above extremality becomes
\begin{equation}
\Delta m_{\rm branes} = (1-\zeta) N_p \tau_p {\frac{(9-p)}{(7-p)}} \,.
\end{equation}
Near extremality we have\footnote{For reasons discussed in \cite{peet}
etc., the cases $p=5,6$ are more problematic to interpret.  We
therefore restrict ourselves from here on to the cases $p\leq 4$.} for
the Hawking temperature
\begin{eqnarray}\label{THzetane}
T_H &=& [2(1-\zeta)]^{\frac{(5-p)}{2(7-p)}} \left(
{\frac{\alpha_p}{16\pi GN_p\tau_p}} \right)^{\frac{1}{(7-p)}}
{\frac{(7-p)}{4\pi}} {\frac{k_p(\lambda_i)}{\rho(\lambda_i)}}
\nonumber\\
&=& [2(1-\zeta)(2\pi)^4]^{\frac{(5-p)}{2(7-p)}}
\left[{\frac{(2\pi\alpha_p)}{(g_{\rm
YM}^2N)\ell_s^4}}\right]^{\frac{1}{(7-p)}} {\frac{(7-p)}{4\pi}}
{\frac{k_p(\lambda_i)}{\rho(\lambda_i)}} \,.
\end{eqnarray}
We can obtain the equation of state for the near-extremal black
D$p$-brane by eliminating $\zeta$ in favour of $T_H$,
\begin{eqnarray}\label{ne_energy}
\Delta m_{\rm branes}(T_H) &=& N^2 \left\{ \gamma_p (2\pi)^2
\left(2\pi\alpha_p\right)^{-\frac{2}{(5-p)}} \right\} \times
\nonumber\\
&& \times\quad \left(g_{\rm{YM}}^2N\right)^{\frac{(p-3)}{(5-p)}}
\left[ {\frac{4\pi}{(7-p)}} {\frac{\rho(\lambda_i)}{k_p(\lambda_i)}}
T_H \right]^{\frac{2(7-p)}{(5-p)}} \,.
\end{eqnarray}
Defining the abbreviation
\begin{equation}
\gamma_p \equiv {\frac{(9-p)}{2(7-p)}} \,,
\end{equation}
we have for the entropy density on the branes
\begin{eqnarray}\label{ne_entropy}
s_{\rm branes}(T_H) &=& N^2 \left\{ {\frac{4\pi}{(7-p)}} (2\pi)^{2}
\left(2\pi\alpha_p\right)^{-\frac{2}{(5-p)}} \right\} \times
\nonumber\\ 
&& \times\quad \left( g_{\rm YM}^2N \right)^{\frac{(p-3)}{(5-p)}} \,
\rho(\lambda_i) \, \left[ {\frac{4\pi}{(7-p)}}
{\frac{\rho(\lambda_i)}{k_p(\lambda_i)}} T_H
\right]^{\frac{(9-p)}{(5-p)}} \,.
\end{eqnarray}
This equation is of central importance; it encodes the equation of
state for the system.

Now we come to the crucial step.  Following the innovation in
\cite{Danielsson}, we actually take the lead for the behaviour of the
strongly coupled field theory on the branes and anti-branes by using
the near-extremal supergravity results.  This is tantamount to using
IMSY duality \cite{itzhaki}.  Here, for general $p$, our proposal to
use IMSY duality for the $p\not=3$ case in the model of type
\cite{Danielsson} is a more nontrivial step than in the $p=3$ case
where the field theory behaved in a simple way.  This assumption
involves some nontrivial physics; like \cite{Danielsson}, we are not
taking into account the lack of a clean decoupling limit in the
brane-antibrane model.  Nonetheless, it is our point of view here that
taking the near-extremal supergravity result seriously for the
strongly coupled field theory on both the set of branes and the set of
antibranes is exactly what we need.

We are particularly interested to investigate this story for the
D0-${\overline{\rm{D}}0}$ case. The reason is that in the $d=0+1$ case
we have the hope of actually checking the above assumption explicitly,
using a strong-coupling simulation. In particular, a significant
motivation for our investigation of the non-conformal ($p\not=3$)
cases in the first place was the result of \cite{kabat1,kabat2} in
which the equation of state was approximately reproduced numerically.
We will begin developing the numerical story for the
D0-${\overline{\rm{D}}0}$ case in the next section, but for now we
work out the analytics.

For the open-string gas on (say) the set of D$p$-branes, we have
\begin{equation}
s_{\rm branes} = {\frac{2\pi}{b_p}} \rho(\lambda_i) \sqrt{N} g_{\rm
YM}^{\frac{(p-3)}{(7-p)}} \left(\Delta m\right)^{\gamma_p} \,,
\end{equation}
where we have defined
\begin{equation}
b_p \equiv {\frac{(7-p)}{2}} \gamma_p^{\gamma_p}
(2\pi)^{\frac{(p-4)}{(7-p)}}
\alpha_p^{\frac{1}{(7-p)}} \,.
\end{equation}

Now, for our model, recalling that we have the field theory on both
the branes {\em and} antibranes, we have
\begin{equation}\label{eTHne}
m_{\rm FT} = (2)N \tau_p + (2)\Delta m(T_{\rm FT}) \,.
\end{equation}
BPS branes carry no macroscopic entropy, so we take the entropy of the
strongly coupled field theory system representing the neutral black
brane to be twice the entropy of the gas on each set of branes,
$s_{\rm FT}(T_{\rm FT}) = (2)s(T_{\rm FT})$.  We find
\begin{equation}\label{SFTrot}
s_{\rm FT} = (2) {\frac{2\pi}{b_p}} \rho(\lambda_i) \sqrt{N} \,
g_{\rm{YM}}^{\frac{(p-3)}{(7-p)}} \, \left( {\frac{m_{\rm
FT}}{(2)}}-N\tau_p\right)^{\gamma_p} \,.
\end{equation}
We now need to know how $\rho(\lambda_i)$ depends on variables of
interest.  We again use the trick of the previous subsection to write
$\lambda_i = 2\pi j_i/s$.  Then
\begin{eqnarray}\label{lambdabp}
\lambda_i\rho(\lambda_i)= \lambda_i \left[\prod_{i=1}^n
(1+\lambda_i^2) \right]^{-{\frac{1}{(7-p)}}} &=& {\frac{b_p}{(2)}}
{\frac{1}{\sqrt{N}}} \left(g_{\rm YM}\right)^{\frac{(3-p)}{(7-p)}} j_i
\left( {\frac{m_{\rm{FT}}}{(2)}} -N\tau_p\right)^{-\gamma_p}
\nonumber\\ &=& {\frac{j_i}{R_{{\rm FT}}(N, m_{\rm FT}, g_{\rm
YM},\tau_p)}} \,,
\end{eqnarray}
where
\begin{equation}\label{RFTdefn}
R_{{\rm FT}}(N,m_{\rm FT},g_{\rm YM},\tau_p) = (2) {\frac{1}{b_p}}
\sqrt{N}g_{\rm YM}^{\frac{(p-3)}{(7-p)}} \left(
{\frac{m_{\rm{FT}}}{(2)}} -N\tau_p\right)^{\gamma_p} \,.
\end{equation}
Referring back to the entropy equation (\ref{SFTrot}), we have for
general $p$ and general angular momenta that
\begin{equation}\label{sFTfnofrho}
s_{\rm FT} = 2\pi \, R_{{\rm FT}}(N,m_{\rm FT},g_{\rm YM},\tau_p)
\, \rho(R_{{\rm FT}}(N,m_{\rm FT},g_{\rm YM},\tau_p),j_i) \,.
\end{equation}

At this stage, we may wonder whether the equations for the physical
parameters $\lambda_i$ in terms of $(j_i,m_{\rm FT},N,g_{\rm
YM},\tau_p)$ can actually be solved explicitly analytically.  For some
$p$, they can.  For other cases, including the $p=0$ case of most
interest to us, however, they cannot.  For now, we will put this issue
aside, and just proceed with the dependence of $\rho$ on $R_{\rm FT}$
and thereby on $(N,m_{\rm FT}, g_{\rm YM},\tau_p, j_i)$ implicit.

\subsection{Comparing field theory and supergravity}
We now take the field theory result of the last subsection and ask
what happens when we optimize with respect to $N$, the number of
branes (and anti-branes).  For simplicity, we first ask how this story
works without angular momenta.  Later we add angular momenta back in.

Optimizing the entropy of the strongly coupled brane-antibrane
theories w.r.t. $N$ gives
\begin{equation}
N = {\frac{m_{\rm FT}}{2\tau_p(1+2\gamma_p)}} \, \quad {\rm i.e.}\quad
m_{\rm FT} = (2N\tau_p) 2\delta_p \,.
\end{equation}
The energy density in the brane gases is then
\begin{equation}\label{egasebranes}
m_{\rm FT} - 2N\tau_p = (2N\tau_p)2\gamma_p \,.
\end{equation}
Substituting these expressions for $N$ and $m_{\rm FT}$ back into the
expression for the entropy, and converting D$p$-brane quantities into
the Newton constant, we obtain
\begin{equation}
s_{\rm FT} = 2^{-\gamma_p} {\frac{2\pi}{a_p}}
G^{\delta_p-1}m_{\rm FT}^{\delta_p} \,.
\end{equation}
Two possible conclusions can be drawn from this.  The first is that
\begin{equation}
m_{\rm FT} = m_{\rm SG} \quad{\rm{and}}\quad s_{\rm
FT}=2^{-\frac{(9-p)}{2(7-p)}} \, s_{\rm SG}
\end{equation}
Alternatively, we can conclude that
\begin{equation}
s_{\rm FT} = s_{\rm SG} \quad{\rm{and}}\quad m_{\rm FT} =
2^{\frac{(9-p)}{2(8-p)}} \, m_{\rm SG} \,.
\end{equation}
The second interpretation has an interesting conclusion. It says that
the field theory energy, which is simply the energy on the D$p$-branes
plus the energy on the ${\overline{{\rm D}p}}$-branes, is not simply
the supergravity energy, but there is a (suitably negative) binding
energy
\begin{equation}
\left| m_{\rm binding} \right| = m_{\rm SG} \left(
2^{\frac{(9-p)}{2(8-p)}} -1 \right) \,.
\end{equation}
We find this interpretation the more attractive one.  Binding energy
can be expected in our model, because of the lack of a {\em clean}
decoupling limit in our system between the open-string and
closed-string modes.

Let us now add back the angular momenta and see if it affects our
exposition of the basic physics of the brane-antibrane model.

Recall that, in the microcanonical ensemble that we are studying, both
$m_{\rm FT}$ and $j_i$ are constants.  This is why we have labelled
$R_{\rm FT}(N$) a function of $N$, which we have to optimize following
the model of $\cite{Danielsson}$.

Regardless of whether the equation for $\lambda_i$ can be solved
explicitly, the entropy depends only on $\rho(R_{\rm FT})$, as in
(\ref{sFTfnofrho}). So we just proceed with $\rho(R_{\rm FT})$
implicit.

Our principle is to maximize the entropy as a function of $N$.  We
have
\begin{eqnarray}\label{maxiboo}
{\frac{1}{s_{\rm FT}}}{\frac{\partial s_{\rm FT}}{\partial N}} &=&
{\frac{1}{R_{\rm FT}}} {\frac{\partial R_{\rm FT}}{\partial N}} +
{\frac{1}{\rho(R_{\rm FT})}} {\frac{\partial\rho}{\partial R_{\rm
FT}}} {\frac{\partial R_{\rm FT}}{\partial N}} \nonumber\\
&=& \left[ {\frac{1}{R_{\rm FT}}} + {\frac{1}{\rho(R_{\rm FT})}}
{\frac{\partial\rho}{\partial R_{\rm FT}}} \right] {\frac{\partial
R_{\rm FT}}{\partial N}} \,.
\end{eqnarray}
Since we do not in general know the analytic dependence of $\rho$ on
$R_{\rm FT}$, we need to find the derivative implicitly.  We have
from (\ref{RFTdefn}) 
\begin{equation}
\lambda_i {\frac{1}{\rho}} {\frac{\partial\rho}{\partial R}} =
-{\frac{j_i}{R^2\rho}} - {\frac{\partial\lambda_i}{\partial R}} \,.
\end{equation}
Using (\ref{sFTfnofrho}), and $\lambda_i/(2\pi)=j_i/s$, we have
$\lambda_i=j_i/(\rho R)$, which gives
\begin{equation}
\lambda_i\left[ {\frac{1}{R_{\rm FT}}} + {\frac{1}{\rho(R_{\rm FT})}}
{\frac{\partial\rho}{\partial R_{\rm FT}}} \right] = -
{\frac{\partial\lambda_i}{\partial R_{\rm FT}}}
\end{equation}
Referring back to (\ref{lambdabp}), it is easy to find $\partial
\lambda_i/\partial R$,
\begin{equation}
{\frac{\left[(7-p)+(5-p)\lambda_i^2 \right]\lambda_i^{6-p}}{
(1+\lambda_i^2) \prod_j(1+\lambda_j^2) }} 
{\frac{\partial\lambda_i}{\partial R_{\rm SG}}} =
-{\frac{(7-p)}{R_{\rm SG}^{8-p}}} j_i^{7-p}
\end{equation}
Therefore the derivative is manifestly negative, as long as $p\leq 4$,
and it is nonzero as long as at least one angular momentum parameter
($j_i$) is turned on.  Consequently, the term in square brackets in
(\ref{maxiboo}) does not vanish, and it is nonsingular.
Therefore, whatever the behaviour of $\rho(R_{\rm FT})$, we are safe
in concluding that the entropy is extremized by demanding that
$\partial R_{\rm FT}(N)/\partial N=0$.  Optimizing $R_{\rm FT}(N)$ we
find that
\begin{equation}
N = {\frac{m_{\rm FT}}{2\tau_p(1+2\gamma_p)}} \,.
\end{equation}
which is exactly what we had for the non-rotating case.

Substituting back this optimal value of $N$ into the field theory
quantity $R_{\rm FT}$, we find that
\begin{equation}
R_{\rm FT}(m_{\rm FT},j_{i\,{\rm FT}},g_{\rm YM},\tau_p)
= R_{\rm SG}(m,j_{i\,{\rm SG}},G) \,,
\end{equation}
where $j_{i\,{\rm FT}}$ are the angular momenta in {\em one} copy of
the strongly coupled field theory, if we make the identifications
\begin{equation}\label{idees}
m_{\rm FT} = m_{\rm SG} 2^{\frac{(9-p)}{2(8-p)}} \,,\quad
j_{i\,{\rm SG}} = j_{i\,{\rm FT}} \,.
\end{equation}
Now, na\"ively we would have expected the total angular momenta of the
neutral supergravity solution to be split in half, shared equally
between the strongly coupled brane and antibrane field theories.  The
fact that the angular momenta and the mass do not match precisely is
not particularly surprising, however, because of the lack of a clean
decoupling limit.  In fact, closed-string modes (whose physics not
included in the brane-antibrane model) might carry both mass and
angular momenta.  We find it intriguing, though, that the
renormalization factors are simply powers of two!

Now, in order to get the entropy to match between the field theory
side and the supergravity side here, we have the freedom to apply
renormalization factors to both the mass and angular momenta.
Alternatively, there is insufficient information to set these
renormalizations factors unambiguously using our analysis. 

Let us then accept the mass and angular momenta renormalizations of
(\ref{idees}).  We then find the remarkable fact that the functional
form of the field theory entropy (\ref{sFTfnofrho}) in terms of $R$ is
identical to the functional form of the entropy on the supergravity
side (\ref{sSGfnofrho}).  With the renormalizations (\ref{idees}), we
see that the $R$'s are the same on both sides.  The strongly coupled
field theory of branes and antibranes therefore reproduces the
supergravity result, up to renormalization factors of two, i.e.
\begin{equation}
s_{\rm FT}\left(m_{\rm FT}=2^{{\frac{(9-p)}{2(8-p)}}} m_{\rm
SG},j_{i\,FT}  \right) = s_{\rm SG}(m_{\rm SG},j_{i\,SG}) \,.
\end{equation}
This is our main analytic result. It shows that, regardless of
rotation or $p$, the entropy of the neutral black brane can be
recovered from the strongly coupled brane-antibrane field theory.  It
is also worth noting that this result {\em cannot} be obtained just by
dimensional analysis.  

Alternatively, we can think of this result as constituting a highly
nontrivial relationship between the thermodynamical properties of
near-extremal black branes and those of neutral black branes.

Before moving on to develop some more physics of the brane-antibrane
model, we may ask what our conclusion about the mass renormalization
may do to other parts of our brane-antibrane-gas model.  We know that
\begin{equation}
T_H^{-1} = {\frac{\partial s_{\rm SG}}{\partial m_{\rm SG}}} \,.
\end{equation}
Therefore if, as we have shown, the entropies match but the masses are
not equal, then the temperature is also affected in the same
proportion as the mass
\begin{equation}
T_{\rm FT} = T_H 2^{\frac{(9-p)}{2(8-p)}} \,.
\end{equation}
(Note that this does not spoil any of our previous assumptions.)  

One might wonder how the unusual thermal properties of neutral black
holes and black branes, such as negative heat capacity, could possibly
be understood from analyzing this D$p$-${\overline{{\rm D}p}}$ field
theory model, which is based on ordinary super-Yang-Mills systems in
various dimensions. The explanation of this point in the context of
the D$p$-${\overline{{\rm D}p}}$ model is simple and was given for the
conformal cases in \cite{Danielsson}.  Upon inspection, one finds that
there is a correlation between the energy in the open string gas and
the contribution to the energy coming from the tension of the branes.
Actually they are proportional to each other. This would give rise to
the following interpretation for the negative specific heat.  Namely,
that since the energy in the gases is proportional to the tension
energy, if we add more energy to the open string gases then we must
create more D$p$-${\overline{{\rm D}p}}$ pairs to maintain the
proportionality.  This requirement to make more D$p$-${\overline{{\rm
D}p}}$ pairs makes the system cool down, meaning that the more massive
the black branes the colder they get.  As we see, the moral of the
story is that `normal' SYM field theories on the worldvolume of the
branes and antibranes behave `abnormally', because the total number of
degrees of freedom controlled by $N$ is not a constant; rather, it is
given thermodynamically by the entropy maximization scheme.

So it is natural and important to ask whether the same kind of
correlations also hold in our particular system of interest, i.e.
D0-${\overline{{\rm D}0}}$, and in an even broader sense, for a
general D$p$-${\overline{{\rm D}p}}$ system.

It is easy to show the linear proportionality (\ref{egasebranes})
between the energy in the gas and the energy contribution from the
brane tension.
Using 
data from the optimization of $N$
\begin{equation}
N\sim {\frac{m_{\rm FT}}{\tau_p}} \,,
\end{equation}
where $m_{\rm FT}= 2N\tau_p + m_{\rm gas}$, gives
\begin{equation}
m_{\rm gas}\sim N\tau_p \,.
\end{equation}
It is intriguing that exactly the same behaviour was observed in the
case of non-dilatonic branes (D3, M2, M5).  However, the way the
result of \cite{Danielsson} was obtained, comparing energy density and
pressure in field theory and supergravity, does not apply to our $p=0$
case of particular interest as there is no pressure on a $0+1$
dimensional worldvolume.

It is important to note that in our entire analysis, the black objects
under study are neutral - nonperturbatively nonextremal.  It is quite
pleasing that the brane-antibrane model precisely explains the entropy
of the neutral black branes and black holes, up to renormalizations of
the mass and angular momenta that we computed.

\subsection{Horizon size}

It is interesting to ask whether the transverse fluctuations of the
D$p$-$\overline{\rm D}p$ system in the microscopic picture can
reproduce the size of the horizon of the corresponding black brane
geometry.  We now do a scaling analysis, not keeping precise numerical
factors.

Before addressing the general-$p$ cases, it is instructive to review
the simplest case $p=3$.  For $N$ {\em near-extremal} D3-branes at
strong coupling, power counting in $N$ and conformal symmetry tell us
that\footnote{The r.m.s. expectation value is normalized with a factor
of $1/N$.}  $\langle {\vec{X}}^2 \rangle_{\rm rms}\sim N T^2$.  Using
the model of \cite{Danielsson} to find the optimal value of $N$ gives
$N\sim m_{\rm FT}/\tau_3\sim m_{\rm gas}/\tau_3$.  Here we have used
the information from the brane-antibrane model both to set the optimal
value of $N$ and to learn that there is roughly the same amount of
energy density in the branes and in the open-string gas on those
branes.  Next, we use the assumption $m_{\rm SG}\sim m_{\rm FT}$, and
the supergravity relationship between the Hawking temperature and the
energy, $m_{\rm SG}\sim r_0^4/G$.  We also recall the assumption that
the brane gas temperature is equal to the antibrane gas temperature,
and both are equal to the Hawking temperature of the black brane.
Collecting these facts, and using the relationship $G\tau_3^2\sim 1$,
then gives a relationship between the supergravity horizon radius and
variables in the field theory.  The last assumption to be used is the
supergravity relationship between the Hawking temperature and horizon
radius $T_H\sim 1/r_0$.  So $\langle {\vec{X}}^2 \rangle_{\rm rms}
\sim r_0^4 (1/r_0)^2 \sim r_0^2$.  Therefore, the r.m.s. extent of the
position fields of the field theory corresponds to the horizon radius.
In other words, we started from a near-extremal gauge theory and ended
up with the size of the horizon of a {\em neutral} black 3-brane.

Now we turn away from the conformal case.  In later sections, we will
be particularly interested in the $p=0$ case, so let us look at it
here in some detail.

In the mean field approximation, it has been shown \cite{KabatLowe}
that the extent of the ground state of the D0-brane theory at low
temperatures is controlled by t'Hooft coupling, {\em not} the
temperature.  In supergravity, on the other hand, the rough estimate
of the size {\em is} temperature dependent.  Therefore, the above
logic that we used for the D3-brane case will not work for the
D0-brane case. In \cite{KabatLowe}, a remedy for this confusing
situation was proposed. Namely, that fast-varying degrees of freedom
are physically inaccessible to a local supergravity observer. The
supergravity probes just cannot resolve high-frequency fluctuations of
order of t'Hooft energies, that are occurring at a ``microscopic''
level in quantum mechanics.  In other words, a supergravity probe
cannot resolve distances sharper than $l_{\rm probe}\sim \beta$. As a
consequence, the supergravity probes simply miss all the quantum
dynamics of the D0-brane dynamics involving energies much higher than
the Hawking temperature.  Therefore, in making a microscopic model to
reproduce supergravity, caution must be used regarding which degrees
of freedom have to be included in the picture.

Practically, the suggestion of \cite{KabatLowe} amounts to imposing a
temperature-dependent cutoff on the spectral density of the transverse
fluctuations, i.e. $X^i$ propagators, that leads to a temperature
dependent $\langle{\vec{X}}^2\rangle_{\rm rms}$ which qualitatively
shows the same behavior as $r^2_0$ does on supergravity side.  

So in what follows we will assume that, if one tries hard from the
gauge theory point of view, the extension of the wavefunction of the
near-extremal geometry should become consistent with supergravity
expectations.  We use this intuition motivated by IMSY in what
follows.

We now apply this thinking that came from studying D0-brane physics to
the general-$p$ case.  We put together a number of ingredients that we
have discussed in this section.  We begin with
\begin{equation}
m_{\rm FT}\sim m_{\rm branes} + m_{\rm
gas} \,,
\end{equation}
and note from the model of \cite{Danielsson} we have 
\begin{equation}
m_{\rm gas}\sim m_{\rm branes} \,.
\end{equation}
(Of course, the same energy density is in the brane gas and antibrane
gas, so for the purposes of scaling we do not need to compute each
contribution separately.)  

From e.g. \cite{itzhaki}, we have for the spatial extent of a {\em
near-extremal} D$p$ or ${\overline{{\rm D}p}}$ geometry,
\begin{equation}
U_0\sim (g_{\rm YM}^2 N)^{1/(5-p)} T^{2/(5-p)} \,,
\end{equation}
where $U_0\equiv r_0/\ell_s^2$.  
(Thinking of $U_0$ as distance and $T$ as energy, we see that this is
exactly the same energy/distance relation found by \cite{peet} for a
$D=10$ supergraviton probe of a D$p$-brane geometry in the decoupling
limit.  The coincidence is not too surprising.)
We now add in dynamical information from the brane-antibrane model,
\begin{equation}
N\sim m_{\rm FT}/\tau_p \,,
\end{equation}
to give
\begin{equation}
r_0^{5-p}\sim G m_{\rm FT} T^{2/(5-p)} \,.
\end{equation}
Lastly, we bring in the (field theory) relationship between
temperature and mass
\begin{equation}
T \sim (Gm_{\rm FT})^{-1/(7-p)} \,.
\end{equation}
Finally, we find 
\begin{equation}
r_0\sim (Gm_{\rm FT})^{1/(7-p)} \,.
\end{equation}
Since $m_{\rm FT}\sim m_{\rm SG}$, this is the same radius as that of
the horizon in the {\em neutral} geometry of interest.  Therefore, the
brane-antibrane model does successfully provide a consistent picture
of the spatial extent of the horizon.

We now turn to checking the consistency of the supergravity
approximation itself.

\subsection{Validity of supergravity}

The black D$p$-branes on which we concentrate here are neutral.  In
order that they can be thought of as bona fide supergravity entities,
we require at a minimum that the curvature at the horizon radius
should be small in string units, to keep $\alpha^\prime$ corrections
small.  In scaling,
\begin{equation}
(Gm)^{1/(7-p)} \gg \ell_s \,.
\end{equation}
Using the equilibrium value of $N$, this becomes
\begin{equation}
(g_{s}^2\ell_s^8N\tau_p)^{1/(7-p)}\gg \ell_s \,,
\end{equation}
which in turn leads to
\begin{equation}
g_{s}N \gg 1 \,.
\end{equation}
Of course, we also require that string loop corrections be under
control
\begin{equation}
g_s \ll 1 \,,
\end{equation}
for the $D=10$ neutral black brane spacetime.

Now, let us recall one important piece of physics from
\cite{Danielsson}, the study of the conformal case.  Even though the
neutral geometry of interest is not studied in the decoupling limit,
the brane and antibrane systems in the microscopic model are actually
taken to be decoupled -- as facts from AdS/CFT are used to describe
the strongly coupled D3-brane theory.  Therefore, at this point, it is
appropriate to check whether the conditions for validity of the $D=10$
near-extremal D$p$-brane geometry are satisfied here also for the
non-conformal cases, as these geometries dictate for us the behaviour
of the strongly coupled field theories on the D$p$ and
${\overline{{\rm D}p}}$.

Let us therefore study the IMSY conditions carefully, to gain
understanding of the brane-antibrane side of the picture.  As
described in \cite{itzhaki}, there is a region in which type II $D=10$
supergravity is valid
\begin{equation}\label{imsycondition}
1\ll g_{\rm eff}^2(U) \ll N^{\frac{4}{(7-p)}} \,,
\end{equation}
where the effective coupling at energy scale $U$ is
\begin{equation}\label{boo1}
g_{\rm eff}^2(U) \sim g_{\rm YM}^2 N U^{p-3} \,.
\end{equation}
Also, the horizon radius for the near-extremal geometry is related to
the energy above extremality $m_{\rm gas}$ and the temperature $T$ by
\begin{equation}\label{boo2}
U_0 \sim (g_{\rm YM}^4 m_{\rm gas})^{\frac{1}{(7-p)}} \sim (g_{\rm
YM}^2N)^{\frac{1}{(5-p)}} T^{\frac{2}{(5-p)}} \,.
\end{equation}

Now, our big neutral black brane will have a low Hawking temperature.
This means that we are not in danger of violating the $\alpha^\prime$
(left-hand) end of the bound (\ref{imsycondition}).  However,
precisely because we are operating at such low temperatures, we may be
concerned about violating the strong-coupling (right-hand) end of the
bound (\ref{imsycondition}).  For example, for $p=0$, we may be
concerned about having to lift up to $D=11$.

It is a satisfying fact that our microscopic brane-antibrane model
remains consistent in the $D=10$ picture.  The essential reason for
this is that, in the brane-antibrane model, $N$ is not an independent
variable.  Rather, $N$ is thermodynamically determined.  Using the
fact that $m_{\rm gas}\sim N\tau_p$ and formul\ae\
(\ref{boo1},\ref{boo2}), and requiring that both ends of the IMSY
bound (\ref{imsycondition}) are respected gives two conditions,
which reduce to
\begin{equation}
g_sN\gg 1 \,, \qquad g_s \ll 1 \,.
\end{equation}
As we saw before, open strings are strongly coupled, but closed
strings are weakly coupled.

Therefore, we do not need to be concerned about departing from the
regime of validity of $D=10$ supergravity, for our systems of strongly
coupled branes and antibranes.

In particular, for the $p=0$ case, we are always in the $D=10$
supergravity regime.  This means, in particular, that we never get to
the $D=11$ supergravity regime.  Therefore, it is not apparent whether
there is any simple relationship between these microscopic models for
$D=10$ neutral black branes and the BFSS matrix theory models, whose
microphysical description is in terms of D0-brane degrees of freedom
representing $D=11$ M theory.

\section{Numerically investigating 
${\rm\bf D0}{-}\overline{\rm\bf D0}$ physics}

For the remainder of this paper we will concentrate on the $p=0$ case,
i.e. the brane-antibrane model of the $D=10$ Schwarzschild black
hole. The reason is that the physics is a quantum mechanics, lending
itself to the possibility of actually computing the behaviour of the
strongly coupled field theory.

\subsection{Scalings, and strategy}

Of course, the idea of doing direct numerical simulations in the
strongly coupled QM is not new.  Kabat et al.~\cite{kabat1} have used
a method called Variational Perturbation Theory (VPT) to check the
IMSY gravity/gauge duality conjectures \cite{itzhaki} for the
decoupling limit of D$p$-brane systems with sixteen supercharges.  In
particular, the entropy of $N$ near extremal D0-branes was computed
using VPT, and was found to match the with supergravity result to the
level of approximation used \cite{kabat2}. The agreement is
impressive.  In particular, the highly nontrivial temperature
dependence of the free energy was obtained (approximately): $\beta
F\propto T^{1.8}$!  Indeed, this could be considered as the first
(approximate) nonperturbative check of any gauge theory/gravity
duality.  The causal structure of spacetime from the point of view of
the gauge theory also has been studied in the context of mean field
Gaussian approximations~\cite{kabat3}.

The next step for our program to understand $D=10$ Schwarzschild black
holes would be to justify the crude brane-antibrane picture drawn in
the last few sections, by computing the microscopic entropy using the
effective quantum description of a system of D0-$\overline{\rm D0}$s.

Now, in the D0-$\overline{\rm D0}$ system we cannot take a clean
decoupling limit, if we expect to keep the open string tachyon and
massless string modes but not massive string modes.  Therefore, our
further progress in developing the brane-antibrane model is to be
thought of as an approximate description, where the massive string
modes are not fully decoupled.  Of course, the same story was true of
the Danielsson et al. work \cite{Danielsson}.

Clearly, there are two different energy scales set by two different
dimensionful coupling constants in our problem: $\alpha'$ and $g_{\rm
YM}^2=g_{s}(2\pi)^{-2}\ell_s^{-3}N$. In gauge theory, everything is
governed by $g_{\rm YM}^2$, while $\alpha'$ controls the mass of the
tachyon.  For the physics of D0-branes alone, dimensional analysis and
power counting in $N$ tells us \cite{kabat1} that a dimensionless
quantity like $\beta F$ can be written as
\begin{equation}
\beta F_{\rm D0} \propto N^2 \mathcal{F} (\frac{T}{(g_{\rm
YM}^2N)^{1/3}}) \,.
\end{equation}
Here, with the tachyon, we have two different scales: the t'Hooft
coupling and $\alpha'$.  Thus, the free energy can be written as
\begin{eqnarray}
\beta F_{{\rm D0}{-}\overline{\rm D0}} \propto N^2 \mathcal{G}
(\frac{T}{(g_{\rm YM}^2N)^{1/3}}, \alpha'(g_{\rm YM}^2N)^{2/3})= N^2
\mathcal{G} (\frac{T}{(g_{\rm YM}^2N)^{1/3}}, (g_{s}N)^{2/3}) \,.
\end{eqnarray}
Therefore, the free energy written in dimensionless units will be a
function of $g_{s}N$.  Of course, it will also depend on the
dimensionless inverse temperature measured in 't Hooft units,
$\tilde\beta$.

In the limit where massive open string excitations can be neglected,
and the string coupling is small, the system is relatively simple.  We
have two copies of the $D0$-brane theory - one for D0-branes and one
$\overline{\rm D0}$-branes - plus a complex tachyon $T$ and a massless
Majorana fermion $\Psi$ coming from D-$\overline{\rm D}$ open
strings.

The strategy is to compute the free energy of this system as a
function of the tachyon classical background field, plus other
quantities at finite inverse temperature $\beta$.  In the field theory
we compute at strong couplings using VPT, and in the closed string
description the $D=10$ supergravity approximation is valid.  Then the
following pieces of information could be read off immediately from the
free energy
\begin{description}
\item[Tachyon static mass] The thermodynamically favourable value for
the tachyon expectation value can be computed by minimizing the
effective action, i.e. the free energy of the system as a function of
the tachyon background.
\item[Sign of the Tachyon dynamical mass] Another important piece of
information encoded in the free energy is the effective mass of the
tachyon which is given by
\begin{equation}
m^2_{T}=\beta\frac{\delta^2F}{\delta \langle T\rangle\delta \langle
T\rangle^{\star}} \,,
\end{equation}
where $\langle T\rangle$ is the tachyon expectation value.  This
effective mass includes the contributions from infinite numbers of
loop diagrams via the Schwinger-Dyson equation.
\item[Magnitude of the Tachyon Dynamical Mass] This quantity is a
measure of the smallness of the tachyon fluctuations.  A large
dynamical mass results in small contributions to the entropy from the
tachyon.
\item[Phase Portrait of the theory] The final goal is to calculate the
phase portrait of the system, as a function of inverse temperature
$\beta$ and $g_sN$. A sign change in $m^2_{T}$ from negative to
positive would be of great interest, since it would signal the tachyon
stabilization phenomenon.  This stabilization would give a
justification for why a gas of D0 and $\overline{\rm D0}$ branes does
not simply annihilate all the way down to a bunch of closed strings.
\end{description}

One note of caution.  The Hagedorn phenomenon in string theory might
impact us here in one place: at temperatures of order of the string
scale, massive open string excitations become important, but we are
not incorporating massive excitations into our dynamics. In order to
have a self-consistent description of the phenomena, therefore, we
will look for a possible finite-temperature tachyon stabilization at a
temperature well below the Hagedorn temperature.

We now turn to describing the technology which we will use to perform
the numerical simulation of the D0-${\overline{{\rm D}0}}$ system.

\subsection{Variational perturbation theory (VPT)}
The basic idea of VPT is simple.  Using VPT to simulate
strong-coupling D0-brane physics was first introduced in
\cite{kabat1,kabat2}, and we review the salient points here for
refernence.  Suppose that the theory of interest has action $S$, and
the aim is to approximate its free energy at {\em strong couplings}.
The idea is to use a free theory, with action $S_0$, with arbitrary
tunable parameters.  The next step is to find values for these
parameters in such a way the free theory is a best fit for the full
interacting theory.

For any arbitrary $S_{0}$ the following identity holds
\begin{equation}
\beta F=\beta F_{0} -\langle e^{-(S-S_{0})}-1\rangle_{0,C} \,,
\end{equation}
where $0$ refers to the free theory and $C$ stands for connected
contributions. Expanding the identity,
\begin{equation}
\beta F=\beta F_{0}+\langle S-S_{0}\rangle_{0}-\frac{1}{2}
\langle(S-S_{0})^2\rangle_{0,C}+ \cdots \,.
\end{equation}
It is important to note that this expansion is {\em not} a
perturbative expansion in the couplings of the original interacting
theory of interest.

Now, if terms to all orders in the above expansion are kept, then
there is of course no dependence on the parameters of the trial free
theory, as it is an identity true for any $S_{0}$.  Taking a practical
approach and terminating the series at any finite order, however, the
series depends on the variational parameters.

The next step is to fix the variational parameters.  Minimizing the
free energy with respect to those parameters (which is equivalent to
requiring the trial free action to satisfy the Schwinger-Dyson
equation) leads to a set of algebraic coupled equations (in general
infinite in number), called ``gap equations".  Solution of the gap
equations yields the variational parameters, which are then
substituted back to obtain the free energy.

This VPT method has in fact been checked explicitly for quantum
mechanical systems where calculations in the full interacting theory
can actually be done exactly, and the above expansion captures the
strong-coupling behaviour accurately; in particular, convergence is
very fast.  Our system of interest here is of course not solvable, so
we need to use the approximate expansion procedure outlined above.

Practically speaking, of course, the infinite set of equations cannot
be solved.  Therefore it is necessary to cut off that infinite set of
equations, and find the common roots of the finite coupled set of
algebraic equations.  Solving this system of equations is, at any
rate, computationally very expensive.

The last step is then to substitute the solution of the gap equations
back into the free energy, to get the sought-after dependence on the
parameters of interest. For us, these parameters will be the
dimensionless inverse temperature $\tilde\beta$, and the open-string
coupling $g_sN$.

As outlined in \cite{kabat1,kabat2}, the VPT method is not
straightforward when dealing with supersymmetry and gauge theory.
Regarding gauge theory.  Proposing a Gaussian theory for a
non-dynamical gauge field in 0+1 dimensions is delicate, as one cannot
simply gauge it away at finite temperature. Gauging away the gauge
field leaves an observable (a Wilson loop), made from the zero mode of
the gauge field encircling the Euclidean time direction, as a remnant.
The Gaussian theory to be used \cite{kabat1} is the one-plaquette
model studied by Gross and Witten~\cite{Gross}.  Taylor-Slavnov
identities, which are consequences of gauge symmetry and relate
various correlation functions, get invalidated by the VPT expansion.
Regarding supersymmetry.  It turns out to be impossible in general to
come up with a Gaussian theory with general variational parameters
respecting supersymmetry, without inclusion of a trial action for
auxiliary fields.  A way to get around the problem is to work with the
off-shell superspace formulation.

On the other hand, the VPT method has some special positive features
as well.  One is that VPT automatically respects 't Hooft counting.
Another is that VPT automatically cures infrared problems arising from
having an infinite moduli space of vacua where D0-branes are far
apart.  The contribution coming from the zero mode sector goes like
$\mathcal{O}(N)$ and is therefore distinguishable from
${\mathcal{O}}(N^2)$ contributions.  

\subsection{Motivating the action}
According to \cite{Kraus} (see also \cite{Tadashi} for a similar
results), the action for a system of D9-$\overline{\rm D9}$ with the
$U(1)$ gauge fields living on its world-volume is given by the
following\footnote{We are using the mostly plus convention for the
metric signature.}
\begin{eqnarray} S&=&-2\tau_{9}\int d^{10}x \,
\exp(-2\pi\alpha^{\prime}T^{\dagger}T)\biggl[ 1 + 8\pi\alpha^{\prime2
}\ln(2)D_{\mu}T^{\dagger}D^{\mu}T + \nonumber\\
&& \qquad\qquad\qquad\qquad\frac{(2\pi\alpha^{\prime})^{2}}{8}
{\overline{F}}_{\mu\nu} {\overline{F}}^{\mu\nu}
+\frac{(2\pi\alpha^{\prime})^{2}}{8} F_{\mu\nu}F^{\mu\nu}
\ldots\biggr] \,,
\end{eqnarray}
where we have used `bars' to denote quantities in the ${\overline{\rm
D0}}$ sector and `no bars' to denote quantities in the D0 sector.  Of
course, T is a complex bi-fundamental gauge field so that
$D_{\mu}T=\partial_{\mu}T-iTA_{\mu}+i{\overline{A}}_{\mu}T$.
Dimensional reduction of this action should give the correct action
for the dynamics of lower dimensional D$p$-${\overline{{\rm D}p}}$
systems including our $p=0$ case.  As usual, we substitute
${\overline{A}}_{i}=(2\pi\alpha^{\prime})^{-1}{\overline{X}}_{i}\,,\,
A_{i}=(2\pi\alpha^{\prime})^{-1}X_{i}$.  Performing field
redefinitions, a nonabelian generalization of this toy approximate
action is
\begin{eqnarray}\label{nonabeliantoy}
S&=&-\int dx^0 {\rm{Tr}}\
\exp\left(-\frac{\pi^2\alpha'}{4\ln2}T^{\dagger}T\right)
\biggl[
\frac{1}{2g_{\rm YM}^2}(\partial_{0}T^{\dagger}\partial^{0}T
+i\partial_{0}T^{\dagger}({\overline{A}}^{0}T-TA^{0})  \nonumber\\
&& +i(A^{0}T^{\dagger}-T^{\dagger}{\overline{A}}^{0})\partial_{0}T
-(A^{0}T^{\dagger}-T^{\dagger}{\overline{A}}^{0})
({\overline{A}}_{0}T-TA_{0}) \nonumber\\
&&
-(X^{i}T^{\dagger}-T^{\dagger}{\overline{X}}^{i})
({\overline{X}}_{i}T-TX_{i}))
+2\tau_{0}\mathbb{I}_{N\times N}\biggr] \nonumber\\
&& + S'_{\rm D0}[X^i,A^0]+S'_{\overline{\rm D0}}
[{\overline{X}}^i,{\overline{A}}^0] \,,
\end{eqnarray}
where we use the usual definition of the 't Hooft coupling, $1/g_{\rm
YM}^2=4\pi^2 g_{s}^{-1}\ell_s^{3}$, and where $S'_{\rm D0}$ and
$S'_{\overline{\rm D0}}$ are the corresponding low-energy actions for
D0 and $\overline{\rm D0}$ respectively.  Obviously, $T$ is in the
$N\times {\overline{N}}$ bifundamental representation.

Now, in order to be able to compute the thermal partition
function, we need to Euclideanize the Minkowskian action. First,
we analytically continue the timelike component of both $U(N)$
gauge fields as well as Minkowskian time direction $x^0$ while
leaving the other fields untouched,
\begin{equation}
x^0=-i\tau \,,\quad
iS_E=S_M \,,\quad iA^{0}_E=A^{0}_M \,,\quad
i{\overline{A}}^{0}_E={\overline{A}}^{0}_M \,.
\end{equation}
Next, we Fourier expand $T$, $A^{0}$, ${\overline{A}}^{0}$, $X^{i}$
and ${\overline{X}}^{i}$ in $\tau$ with a periodicity $\beta$ given by
the the inverse temperature.  Since all these variables are bosonic,
they have periodic boundary conditions,
\begin{equation}
\left\{ A_{0}(\tau) , {\overline{A}}_{0}(\tau), X^i(\tau) ,
{\overline{X}}^i(\tau) , T(\tau) \right\}
=\frac{1}{\sqrt{\beta}}\sum_{\ell\in\mathbb{Z}} \left\{ A_{0\ell} ,
{\overline{A}}_{0\ell} , X^i_{\ell} , {\overline{X}}^i_{\ell} , T_\ell
\right\} e^{\frac{-2\pi i \ell}{\beta}\tau} \,.
\end{equation}
Writing the action in terms of Fourier modes is straightforward.  We
take the above expansions and perform the Euclidean time integral,
using the following {\em variational} correlators in Wick contractions
of Feynman diagrams contributing to the free energy (in our case, the
amplitude $\langle S_{II E}-S_{0 E}\rangle_{0}$,)
\begin{eqnarray}
\langle X^{i}_{\ell AB} X^{j}_{m CD}\rangle_{0} &=&\sigma_{\alpha
\ell}^2 \delta^{i j}\delta_{\ell+m}\delta_{AD}\delta_{BC}\quad\quad
i,j=1,2 \,,\nonumber\\ 
\langle X^{a}_{\ell AB} X^{b}_{m CD}\rangle_{0} &=&\Delta_{\alpha l}^2
\delta^{ab}\delta_{\ell+m}\delta_{AD}\delta_{BC}\quad\quad a,b=3..9
\,,\nonumber\\
\langle A_{00AB} A_{00CD}\rangle_{0} &=&
\rho_{0}^{2} \delta_{AD}\delta_{BC} \,,\nonumber\\
\langle T^{\dagger}_{\ell AB} T_{m
CD}\rangle_{0}&=&\xi_{\ell}^2\delta_{\ell m}\delta_{AD}\delta_{BC}
\,.
\end{eqnarray}
and similarly for the barred variables.  The need for separation of
transverse scalar directions comes from the limitations of the
original superspace formulation \cite{kabat2} of the D0-brane
supersymmetric matrix quantum mechanics. Namely, for each copy of the
QM, there are 2 scalars $X^{i}$ in the gauge multiplet and $7$ other
$\phi^{a}$ scalars coming from scalar multiplet.  In other words, the
original $SO(9)$ $R$-symmetry is broken to $SO(2) \times\ G_2$
($\phi^{a}$ are a $\textbf{7}$ of $G_{2}$).

The numerical simulation of this action, which is an approximation
to the system of D0-branes, anti-D0-branes and the tachyon sector
where both copies of the D0-brane theory are supersymmetric is
computationally very expensive, partly because of the need to find
the solution to hundreds of nonlinear equations. In addition, we
have to cover mapping out of the phase portrait as a function of
$\beta$ and $g_sN$.  We therefore report on some initial results
in a more stripped-down version of our model.

\subsection{The toy model}
In this section, we look at a ``toy'' model which consists of
large-$N$ bosonic matrix quantum mechanics (representing the D0-brane
and anti-D0-brane theories) plus our bi-fundamental charged tachyon.

This toy model is obviously a good approximation to the original
problem in the high-temperature limit where the Euclidean time circle
is so small that nonzero thermal KK modes are heavy. In this limit,
there is no antiperiodic fermion left after reduction to $0+0$
dimensions. Of course, we are interested in low temperatures where the
supergravity description is valid and quantum mechanics is strongly
coupled.

There are reasons to believe why this toy model could demonstrate some
generic behaviors of interest to us, which will continue to be true
even in the case where we have two copies of {\em supersymmetric}
matrix quantum mechanics. It is well known \cite{Polchinski} that the
extent of the ground state of a system of D0-branes, defined in terms
of fluctuations of transverse scalars, is controlled by the 't Hooft
coupling. It is intuitively plausible to expect that, at strong
coupling, the tachyon could acquire a large induced mass through its
coupling to these large transverse scalar fluctuations.  As has been
observed in \cite{kabat1}, both bosonic matrix quantum mechanics and
its supersymmetric version have large transverse fluctuations.  So if
one takes this intuition seriously, as far as the dynamics of the
tachyon is concerned in the low energy approximation, even the bosonic
version could be a good approximation to the right
physics\footnote{Although irrelevant to the above discussion, the
difference is in the actual value of the ground state energy, which is
-- for each copy -- of course zero in the supersymmetric case but
large and positive in the bosonic case.}

In the model of Danielsson et al \cite{Danielsson} which motivated our
work here, the dynamical assumption is that the tachyon field is
stabilized by finite-temperature strong-coupling effects about $T=0$.
We therefore expand $T= 0 + t$, where $t$ is the fluctuation part, and
plan to show the consistency of this assumption {\em a posteriori} by
showing that the mass is indeed positive -- and large\footnote{We have
ignored the higher powers in the expansion of the exponential
prefactor in (\ref{nonabeliantoy}) sitting in front of the Lagrangian;
this will be justified {\rm a posteriori} by the large-mass finding.}
-- at strong coupling.  Therefore, the terms relevant to our
discussion which are quadratic in the tachyon fluctuations $t$ can be
written
\begin{eqnarray}
S_{E}&=&\frac{1}{g^2_{\rm YM}}\int_{0}^{\beta}
d\tau(\frac{1}{2}{\rm{Tr}} D_{\tau}X^{i}D_{\tau}X^{i} -\frac{1}{4}
{\rm{Tr}}[X^{i},X^{j}][X^{i},X^{j}])\nonumber \\
&& +\frac{1}{g^2_{\rm YM}}\int_{0}^{\beta} d\tau(\frac{1}{2}{\rm{Tr}}
D_{\tau}{\overline{X}}^{i}D_{\tau}{\overline{X}}^{i} -
\frac{1}{4}{\rm{Tr}}[{\overline{X}}^{i},{\overline{X}}^{j}]
[{\overline{X}}^{i},{\overline{X}}^{j}]) \nonumber\\
&& +\int_{0}^{\beta} d\tau {\rm{Tr}}[\frac{1}{2g^2_{\rm
YM}}
\biggl(\partial_{\tau}t^{\dagger}\partial_{\tau}t
+A^{0}t^{\dagger}tA^{0}+t^{\dagger}{\overline{A}}^{0}
{\overline{A}}^{0}t +X^{i}t^{\dagger}tX^{i} +
t^{\dagger}{\overline{X}}^{i}{\overline{X}}^{i}t
\nonumber\\ 
&&
- \partial_0 t^\dagger ({\overline{A}}^0 t - t A^0) + (A^0
t^\dagger-t^\dagger {\overline{A}}^0) \partial_0 t \biggr)
+2\tau_{0}\mathbb{I}_{N\times
N}-\frac{2\tau_0\pi^2\alpha'}{4\ln2}t^{\dagger}t\biggr] \,.
\end{eqnarray}
Morally, we are going to treat the tachyon as background.  By this we
simply mean that we are interested in finding the free energy as a
function[al] of the tachyon, after integrating out the other fields in
the problem -- nonperturbatively using VPT.  For this reason, no trial
action for the tachyon is introduced; the VPT techniques are used for
the fields other than the tachyon.

Using VPT, we have
\begin{equation}
\beta F=\beta F_{0} + \beta {\overline{F}}_{0}+\langle S_{\rm
D0}+S_{\overline{\rm D0}} +
S_{T}-S_{0}\rangle_{0}-\frac{1}{2}\langle(S_{III})^2\rangle_{0,C}
\quad (+\ldots) \,,
\end{equation}
where $S_T$ is the tachyon contribution to the action, and the
subscript zero refers to the contribution coming from the trial free
action for the other fields.

As a first step, we can neglect the terms in the action that are cubic
in the tachyon, as far as the VPT expansion is concerned.  There is a
good reason why this is justified.  Namely, because the cubic
contribution to the free energy enters at the order of
$\mathcal{O}(t^4)$, which is small at small-$t$. 

Following the story for a single clump of D0-branes \cite{kabat1}, we
put in the ghost field from gauge-fixing, whose Fourier modes are
denoted $s_{\ell\not=0}$ (obviously, for physical reasons there is no
zero mode).  We also use equations from pages 25-27 of that work; to
save space here we only write here what changes when we introduce our
tachyon field.  

We also introduce convenient dimensionless units. These are defined
such that in term of them one gets an overall factor of $N^2$ in the
free energy, and the rest is just a function of dimensionless
quantities and $g_{s}N$.  We define newly dimensionless variables 
with tildes; we have
\begin{equation}
\tilde{\beta}=\beta {f^{1/3}} \,,\quad
\tilde{\rho}_{0}^2=\frac{N\rho_{0}^2}{f^{1/3}} \,,\quad
\tilde{\sigma}_{l}^2=\frac{N\sigma_{l}^2}{f^{1/3}} \,,\quad
{\tilde{t}}_\ell^\dagger {\tilde{t}}_\ell=\frac{N t^\dagger_\ell
t_\ell}{f^{1/3}} \,,
\end{equation}
and similarly for the barred variables.  Here $f=(g_{\rm YM}^2
N)^{1/3}$.  Note that in this bosonic model we do not have to treat
the $X^i,{\overline{X}}^i$ fields for $i=1,2$ differently than for
$i=3\ldots 9$ because there is no need to worry about maintaining
explicit supersymmetry; therefore we do not have $\Delta_\ell$ and we
just use $\sigma_\ell$ to represent correlators of all nine position
fields.  Note that we are measuring the tachyon expectation value in
string units.  

Henceforth, we drop the tildes for notational simplicity.  Then we
obtain the following expression for the free energy as a function of
the tachyon
\begin{eqnarray}
\beta F&=&\beta F(\lambda)_{\square}+\beta
F({\overline{\lambda}})_{\square} +
\frac{N}{\lambda}\langle
{\rm{Tr}}(U+U^{\dagger})\rangle_{\square}+
\frac{N}{{\overline{\lambda}}}\langle
{\rm{Tr}}({\overline{U}}+{\overline{U}}^{\dagger})\rangle_{\square}
\nonumber \\
&&
-\frac{9N^2}{2}\sum_{\ell}\log\sigma_\ell^2
-\frac{9N^2}{2}\sum_{\ell}\log{\overline{\sigma}}_\ell^{2}
+N^2\sum_{\ell\neq 0}\log s_{\ell}
+N^2\sum_{\ell\neq 0}\log {\overline{s}}_{\ell}\nonumber\\  &&+
\frac{9N^2}{2}\sum_{\ell}((\frac{2\pi l}{\beta})^2
\sigma_{\ell}^{2}-1) +\frac{9N^2}{2}\sum_{\ell}((\frac{2\pi
l}{\beta})^2{\overline{\sigma}}_{\ell}^{2}-1) \nonumber\\
&& +N^2\sum_{\ell\neq 0}((\frac{2\pi
l}{\beta})^2s_{\ell}+1)+N^2\sum_{\ell\neq 0}((\frac{2\pi
l}{\beta})^2{\overline{s}}_{\ell}+1) +
\frac{9N^2}{\beta}\rho_{0}^{2}\sum_{\ell}\sigma_{\ell}^{2} \nonumber\\
&&
+ \frac{9N^2}{\beta}{\overline{\rho}}_{0}^{2}
\sum_{\ell}{\overline{\sigma}}_{\ell}^{2}
+\frac{36N^2}{\beta}(\sum_{\ell}\sigma_{\ell}^{2})^2 +
\frac{36N^2}{\beta}(\sum_{\ell}{\overline{\sigma}}_{\ell}^{2})^2
\nonumber\\
&& +N^2\sum_{\ell}\frac{1}{2}(\frac{2\pi l}{\beta})^2
{\rm{Tr}}\frac{t^{\dagger}_{\ell}t_{\ell}}{N^2}+
\frac{N^2}{2\beta}(\rho^{2}_{0} + {\overline{\rho}}^{2}_{0})
\sum_{\ell}{\rm{Tr}}\frac{t^{\dagger}_{\ell}t_{\ell}}{N^2} \nonumber\\
&&+ \frac{9N^2}{2\beta}(\sum_{\ell}\sigma^{2}_{\ell} +
\sum_{\ell}{\overline{\sigma}}^{2}_{\ell})
\sum_{\ell}{\rm{Tr}}\frac{t^{\dagger}_{\ell}t_{\ell}}{N^2}
\nonumber\\ &&
-\frac{(2\pi)^{4/3}N^2}{8\ln2(g_{s}N)^{2/3}}\sum_{\ell}
{\rm{Tr}}\frac{t^{\dagger}_{\ell}t_{\ell}}{N^2} +
\frac{2(2\pi)^{2/3}N^2\beta}{(g_{s}N)^{4/3}} \,.
\end{eqnarray}

In what follows, we use a natural ansatz for the propagators
\begin{equation}
\sigma^{2}_{\ell}= \left( {(\frac{2\pi \ell}{\beta})^2+m^{2}}
\right)^{-1} \,,
\end{equation}
and similarly for the barred variables.

The gap equations are easily obtained. We have (and similarly for
barred variables)
\begin{equation}
m^{2}=\frac{2\rho^{2}_{0}}{\beta} +
\frac{16}{\beta}\sum_{\ell}\sigma^{2}_{\ell} +
\frac{1}{\beta}\sum_{\ell}\frac{t^{\dagger}_{\ell}t_{\ell}}{N^2} \,,
\end{equation}
where $\rho_0$ can be explicitly obtained in terms of $\lambda$, the
coupling in the one-plaquette action, because the one-plaquette theory
is soluble in the large-$N$ limit.

For $\lambda\leq 2$,
\begin{equation}
-1+\frac{2}{\beta}\left(\frac{18}{\beta}\sum_{\ell}\sigma^{2}_\ell +
\frac{1}{\beta}\sum_{\ell}{\rm{Tr}}
\frac{t^{\dagger}_{\ell}t_{\ell}}{N^2}\right)
\left[1-(1-\frac{2}{\lambda})\log(1-\frac{\lambda}{2})\right]=0 \,,
\end{equation}
while for $\lambda\geq 2$, 
\begin{equation}
\lambda =\beta/\left(\frac{18}{\beta}\sum_{\ell}\sigma^{2}_\ell +
\frac{1}{\beta}
\sum_{\ell}{\rm{Tr}}\frac{t^{\dagger}_{\ell}t_{\ell}}{N^2}\right) \,.
\end{equation}
We will solve this system of highly nonlinear, coupled equations.  The
strategy will be to treat the tachyon as a background.  Solving the
gap equations, and substituting back the solution as a function of the
tachyon, then gives the free energy as a function of the (background)
tachyon.  Given this effective action for the tachyon, it is then
straightforward to read off the tachyon effective mass. The plan will
be to work out the two-dimensional ``phase portrait'' of the system
(with coordinates $g_{s}N $ and $\beta$), and find out whether any
tachyon stabilization is possible. By this we simply mean that there
are points in this two-dimensional portrait where the effective
tachyon mass is positive.

Before we move to displaying our numerical results, a number of
remarks are in order.

\begin{itemize}

\item As we saw earlier, in the supergravity limit in which we are
interested, $g_{s}N \gg 1$.  By inspection of the equations it might
then seem that, in the units we use, the wrong-sign mass term for the
tachyon is suppressed and we are done!  This quick-and-dirty reasoning
is, however, too na\"ive. In fact, in order not to excite any massive
open string state, the temperature must be small in string units.  In
turn, this implies that
\begin{equation}
\beta \gg (g_{s}N)^{1/3} \gg 1 \,,
\end{equation}
where $\beta$ is the dimensionless inverse temperature. With such low
dimensionless temperatures, it becomes more and more difficult to
overcome the wrong-sign mass term through the tachyon couplings to the
transverse coordinates $X^{i},{\overline{X}}^i$ and the world-volume
gauge fields $A_{0},{\overline{A}}_0$. 

\item It is crucial to remember that all variational parameters are
{\em implicit} functions of the tachyon $t$.  So it is certainly not
true that we can simply read off the effective tachyon mass from the
the coefficients of the quadratic tachyon fluctuation in the
expression for the free energy.  For example, terms like
$(N^2/\beta)\rho^{2}_{0}\sum_{\ell}\sigma^{2}_{\ell}$, which do not
have any {\em explicit} background tachyon dependence, are becoming
tachyon-dependent through the gap equations.  All such terms make
contributions to the effective tachyon mass, as one expands the
variational parameters around $\sum_{\ell}{\rm{Tr}}
t^{\dagger}_{\ell}t_{\ell}=0$. (Note that these expansions exist,
since the solution to the gap equations is differentiable with respect
to $t$ in the vicinity of $\sum_{\ell}{\rm{Tr}}
t^{\dagger}_{\ell}t_{\ell}=0$.)

\item Obviously, it is always possible to stabilize the tachyon at
weak couplings, but the temperatures needed for this are above the
Hagedorn temperature. It is easy to see this in perturbation theory,
partly because at such high temperatures the D0 QM is perturbative.
Perturbatively, to stabilize the tachyon, it is sufficient to have
\begin{equation}
T>\frac{1}{(g_{s}N)^{4/3}\ell_s} \gg \frac{1}{\ell_s} \,,
\end{equation}
for small $g_{s}N$.

\end{itemize}

In our investigations, we use the Metropolis algorithm.  This is a
fast method generally; it really comes into its own for numerical
investigations of the supersymmetric model, the numerical results from
which we hope to report on in the near future.  We summarize some
important features of the Metropolis algorithm in the Appendix.

\subsection{Numerical results}
We are interested in the sign of the mass of the tachyon zero mode,
for each point in the two-dimensional parameter space, coordinatized
by $g_sN$ and $\beta$ which is the dimensionless inverse temperature
in 't Hooft units.

We now plot our results for the phase portrait of the toy model
theory.  Inverse temperature, in dimensionless 't Hooft units, is on
the $y$-axis, while the open-string coupling $g_s N$ on the $x$-axis.
(As a reminder, the parameter $g_s N$ tells us the bare negative
mass-squared of the tachyon.)  Points at which the tachyon mass
squared is positive are indicated by a cross, and those at which it is
negative are indicated by a circle.  The green curve delineates where
the Hagedorn phenomenon is important; points above the curve are below
the Hagedorn temperature.  The red straight line delineates the region
where the gauged QM theory has a strong effective coupling (set by the
inverse temperature and 't Hooft coupling); the temperatures where the
quantum mechanical theory is strongly coupled lie above the red line.

For clarity, we separate the regimes of smaller $g_s N $ from the
regime of larger $g_s N $.  

\begin{figure}
\begin{center}
\epsfysize=3truein
\epsffile{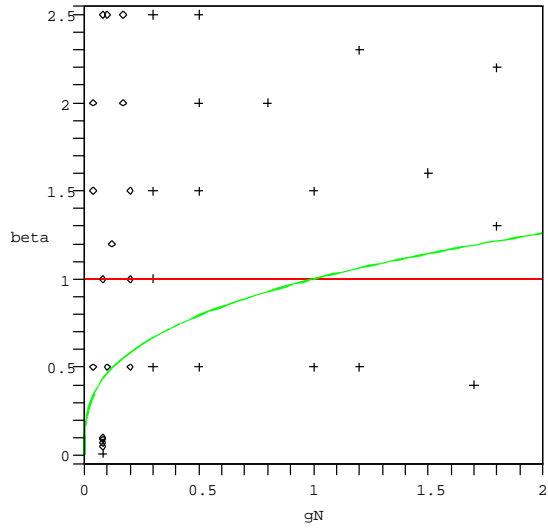}
\end{center}
\caption{Plot of results for the $g_sN\leq 2$ section of the phase
portrait.  }
\label{fig:fig1}
\end{figure}

\begin{figure}
\begin{center}
\epsfysize=3truein \epsffile{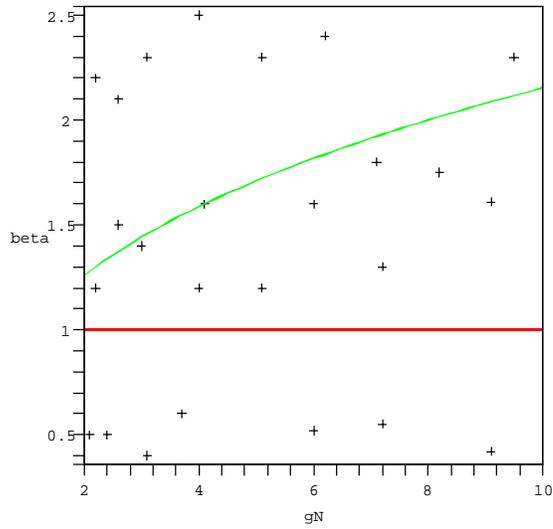}
\end{center}
\caption{Plot of results for the $g_sN \geq 2$ section of the phase
portrait.  }
\label{fig:fig2}
\end{figure}

As we expected for tachyon stabilization, at small $g_sN$, one has to
go to very high temperatures; this is clear from Figure
\ref{fig:fig1}.  As we see for large $g_sN$, on the other hand, even
for low temperatures compared to the Hagedorn temperature, there is a
possibility of tachyon stabilization.

\section{Discussion}

In this paper, following the lines of \cite{Danielsson}, we have
examined the validity of the brane-antibrane model for neutral black
branes of various dimensionality.  What we have found is that this
D$p$-${\overline{{\rm D}p}}$ picture works for arbitrary
$p$. Interestingly, this is true even for the ``dilatonic'' branes,
where the supergravity entropy cannot be written in terms of the
entropy of a free world-volume gauge theory.  In the non-conformal
cases, using IMSY duality proves to be working, relating strongly
coupled gauge theories to the corresponding near-extremal backgrounds.
We were able to show that, even after adding general angular momenta
quantum numbers, the brane-antibrane model is able to reproduce the
supergravity entropy.  The exact agreement in the entropy between the
microscopic brane-antibrane side and the supergravity side comes at
the expense of a renormalization of the mass by a $p$-dependent power
of two, which we interpret as a binding energy, and a renormalization
of the angular momenta by a simple factor of two.  We also find that
the extent of the wavefunction of the $Dp-\overline{Dp}$ system
exhibits the same scaling with the mass as does the radius of the
horizon of the corresponding black brane.

In the original proposal of \cite{Danielsson}, the authors argued that
the tachyon stabilization at finite temperature and strong coupling
could lead to a reappearance of open string degrees of freedom.  Even
at the field theory level, questions regarding strong coupling
dynamics are often hard to answer, but in the special case of
D0-${\overline{{\rm D}0}}$ there is a possibility to address some of
the aspects in the context of a simple toy model following the lines
of \cite{kabat1,kabat2}.  Using techniques developed there, we found
the effective action for the tachyon numerically, in the context of
the toy model. We have seen signals of tachyon stabilization in this
toy model, at strong open string couplings and low temperatures
compared to the string scale.  This might open a window to see clearly
why it is possible to have a long-lived state of D0-branes and
$\overline{{\rm D}0}$-branes, without decaying into closed strings at
couplings of order one and low temperatures.

Our investigation here was for the case of neutral black branes, where
the numbers of microscopic constituent branes $N$ and antibranes
${\overline{N}}$ are equal.  A natural extension of our work here will
be to consider the case where the $D=10$ black branes carry $p$-brane
charge $Q_p=N-{\overline{N}}$.  We hope to report on this story in the
future\footnote{The $p=3$ case with charge was of course previously
considered in \cite{Danielsson,Guijosa}.  It will be interesting to
check whether this agreement holds up also in the {\em non}-conformal
cases.  We thank Alberto Guijosa for a discussion in this regard.}.

Inspired by numerical investigations, we might speculate about the
possible shape of the potential\footnote{at least for $|T|$ in the
vicinity on $T=0$} in this model, say for the simplest case of a
diagonal tachyon condensate at finite temperatures
\begin{equation}
\beta F \propto e^{-\gamma |T|^2}(1+\lambda(N,g_{YM},\beta)
|T|^2+\ldots)
\end{equation}
where $\lambda > \gamma$. This potential has one minimum and two
symmetrically located maxima. The minimum would correspond to the
tachyon stabilization, i.e. the tachyon particle is classically
stable\footnote{The kinetic term probably is not canonically
normalized in the original field.}, even though this vacuum would be
unstable due to quantum and thermal fluctuations and the finiteness of
the height of the barrier. The tachyon can be expected to decay
eventually into closed string radiation. In other words, this
temporary and approximate decoupling between the open string modes and
the closed strings degrees of freedom -- which has apparently
manifested itself in our ability to compute the entropy of the
closed-string background out of Yang-Mills degrees of freedom in the
context of brane-antibrane model -- would not last forever, and the
tachyon particle ground state would have a finite lifetime. If this
picture corresponds well to the true stringy physics, then it would be
plausible to think of the unstable maxima discussed above as being the
unstable state of a D0-$\overline{{\rm D}0}$ pair created from the
energy in the gas living on the worldvolume of the clump of D0s and
$\overline{{\rm D}0}$s, due to the quantum and thermal fluctuations.
This interpretation also leads to an estimate for the
lifetime. Climbing up the hill to create such a pair requires energies
at least of the order of $1/g_s$, so the whole tunnelling process is
suppressed by $e^{-1/g_{s}}$.

\section*{Acknowledgements}

We would like to thank Joel Giedt, Martin Kruczenski, Erich Poppitz,
and Lenny Susskind for helpful discussions.  We would especially like
to thank Dan Kabat for substantive helpful conversations, particularly
about the numerical collaborations \cite{kabat1,kabat2}.  We also
thank Kaveh Khodjasteh and Johannes Martin for assistance with
computer resources.
 
\section*{Appendix}

The problem of finding the common solution to a set of nonlinear
coupled algebraic equation can be though of a minimization process.
To see this, consider the collection of the equations to be solved,
$\mathcal{G}=\{f_{1}=0,...,f_{n}=0 \}$.  Clearly, a quadratic form
$\mathcal{H}$ can be made out of this, viz.  $ \mathcal{H}=\sum_{i}
f^2_{i} $.  If there is a solution to $\mathcal{G}$, this solution
would be the global minimum of $\mathcal{H}$ i.e.,$
f_{1}=0,...,f_{n}=0 $.

To solve this nonlinear system of equations, we use the Monte Carlo
local update method (``Metropolis'' algorithm). This method is one of
the stochastic search methods widely used for minimization of
multivariable functions.

This method does not get trapped in local minima, because of a smart
choice of control variables. The way this method accomplishes the goal
is that, by introducing an unphysical temperature, thermal
fluctuations of the configuration vector are allowed to happen in all
possible directions in the configuration space - even in the ``wrong
directions'' (which increase the Hamiltonian).  By decreasing the
temperature very slowly, one lets the vector find the global minimum
(or minima). For a vector trapped in a local minimum, there would be a
certain probability to climb up the barrier because of the thermal
fluctuations, and so the vector could find its way all the way down to
the global minimum subject to a large number of iterations.  This
story is actually very similar to the physics of the annealing
process, and it is sometimes called the ``stimulated annealing''
method.  With this method, a vector stores the starting point
configuration. This configuration is changed randomly to generate new
child configurations. A new configuration gets accepted with the
probability $1$ if it has a smaller Hamiltonian than its parent, but
if it has not then it would be accepted with the following
Boltzmann-like probability $ P_{\rm acceptance}\sim e^{-\beta_{\rm
fake}\Delta \mathcal{H}}$, where $\beta_{\rm fake}$ is a unphysical
parameter acting like an inverse temperature, and $\Delta \mathcal{H}$
is the energy difference between the child and parent
configurations. After many iterations, and at the same time lowering
the temperature slowly, the configuration vector stabilizes on the
one(s) which minimize(s) the $\mathcal{H}$.

The Metropolis algorithm has strong advantages over the Newton-Raphson
(NR) method, especially when the dimensionality of the configuration
space is big.  Under such circumstances, the NR method becomes very
inefficient, and its starting point dependence can be prohibitively
large.  Practically, this makes it very hard to reach the solution in
reasonable computer time.  The great advantage of the Metropolis
method is the fact that it is almost independent of starting
point.\footnote{Of course, any configuration space could be composed
of disconnected pieces, such that one can not get to any arbitrary
point by starting from another arbitrary point and moving through the
potential.}  Another advantage of Metropolis is that it naturally
avoids getting stuck in local minima, a problem to which NR often
succumbs.


\end{document}